\documentstyle[12pt,epsfig]{article}
\begin{document}

\begin{center}
{\large{\bf Possible Origin of Clusters}}
\begin{center}
{\large{\bf in Ultra-High-Energy Cosmic Rays
}}
\end{center}
\vspace*{5mm}

\bf{A.V. Uryson}

\vspace*{3mm}

{\it  Lebedev Physics Institute, 117924, Moscow, Russia} 
\end{center}

\small{E--mail: uryson@sci.lebedev.ru}

\vspace*{5mm}

{\small{
\centerline{\bf Abstract}
We estimate the detection rate of ultrahigh-energy 
cosmic rays on ground-based arrays by assuming that 
the cosmic-ray sources are active galactic nuclei. 
We analyze the cases of detection of clusters, 
several particles that arrived, within the error limits, 
from the same area of the sky. The adopted model is shown 
to explain the detection rate of clusters on the AGASA array. 

\vspace*{3mm}

{\bf 1.~~ Introduction}

\vspace*{2mm}

The origin of ultra-high-energy cosmic rays (UHECRs), 
$E>4\times 10^{19}$ eV, has not yet been elucidated. 
In our opinion, the sources of UHECRs are active galactic 
nuclei (AGNs). This is an old hypothesis that was 
discussed in the late 1960s and with which the AGN 
space density estimates [1](Berezinskii et al. 1990) 
were consistent. The identification of CR sources performed 
in [2-6] has shown 
that Seyfert nuclei with redshifts $z<0.01$ and 
Blue Lacertae objects (BL Lac) are possible sources of UHECRs. 

On their way from the source to the Earth, UHECRs are 
deflected by magnetic fields in the intergalactic 
space and in the Galaxy. When identifying the sources, 
we assumed that CRs are deflected by the intergalactic 
magnetic fields through no more than $9^0$. 
The intergalactic magnetic fields are $B<10^{-9}$ G [7],
and this condition is satisfied 
for the particles arrived from distances of $\sim 50$ Mpc, 
i.e., emitted by Seyfert nuclei with $z<0.01$ 
[8]. BL Lac objects are hundreds 
of megaparsecs away, but even for such distant 
sources, the deflections may not exceed a 
few degrees if the filamentary structure of the 
intergalactic magnetic fields is taken into 
account [9]. 
The field in the Galaxy has regular and irregular components;
the field strength is $B\sim 10^{-6}$ G. 
In the disk, the field is regular in the 
spiral arms and directed
along them; in the halo, the regular component 
is perpendicular to the disk [7]. 
The deflections of particles in the regular field 
depend on the particle arrival direction and can be 
negligible. In the irregular field, the deflections 
do not exceed $1^0$. Therefore, our identification 
is valid for CRs arrived from fairly high Galactic 
latitudes (in all our identification works, 
we selected UHECRs depending on the Galactic 
latitude of their arrival).

Our identification of sources was based on a 
statistical analysis. The major CR sources can 
be revealed in this way, but it is dificult to 
exclude other hypotheses. For instance, there 
are UHECR particles that arrived from areas of 
the sky where there are neither Seyfert nuclei 
with $z<0.01$ nor BL Lac objects. This may be 
because the catalogs of objects are incomplete; 
they do not contain all objects of a given type, 
and, as a result, some particles arrive from 
''empty'' areas of the sky. (For this reason, 
the sources can be identified only statistically.) 
However, a different
explanation is possible: the statistical analysis 
has revealed the major CR sources, but there are 
also other, less efficient or rarer sources. CRs in 
the arrival direction of which no object of the 
''major'' type is observed originate from these 
''minor'' sources. To reliably establish the CR 
sources, we must find out whether conditions for 
UHECR acceleration exist in the identified objects 
and compare the model predictions with experimental data.

The particle acceleration in the identified objects 
was considered in [8, 10-12]. 
It was shown that conditions for particle acceleration 
to energies of $10^{21}$ eV exist in the identified 
objects. The CR spectra near the Earth obtained in 
the model under consideration agree, within the 
error limits, with the measured spectra [13].

Here, we analyze the clusters detected in UHECRs, 
groups of particles with coincident (within the 
error limits) arrival directions, and consider 
how the existence of clusters can be explained in our model.

\vspace*{3mm}

{\bf 2.~~ The UHECR detection rate from a single source}

\vspace*{3mm}

Let us estimate the detection rate of UHECRs emitted 
by a single source on ground-based arrays with an area 
$S\approx 10$ $\mbox{km}^2$ (the Yakutsk, Haverah 
Park, Akeno, and Volcano Ranch arrays have approximately 
the same area) and $S\approx 100,3000$ $\mbox{km}^2$ 
(the areas of the AGASA and Pierre Auger arrays). 

Let us first consider Seyfert nuclei. To provide the 
observed intensity of CRs with energies 
$E >5\times 10^{19}$ eV, 
$I(E)\approx 10^{39}-10^{40}\,(\mbox{cm}^2\,\mbox{s}\,\,\mbox{sr}\,\mbox{eV})^{-1}$ 
[14], a Seyfert nucleus 
spends an estimated power of 
$L_S\approx 10^{39}-10^{40}\,\mbox{erg}\,\,\mbox{s}^{-1}$ 
on the emission of UHECRs [4,8]. 

This estimate was obtained by assuming that the 
particles are emitted by the source isotropically. 
However, in our acceleration model [8, 12], 
the particles are accelerated in 
a jet and are emitted in a directed beam with an 
opening half-angle of 
$\theta\le 6.5\times 10^{-4}$. Taking into account 
the possible conversion of the accelerated particles 
from the charged into the neutral state and back [15], 
we find that 
UHECRs are emitted in a cone with an angle of $20^0$. 
It thus follows that the sources of the detected CRs 
are Seyfert nuclei oriented in such a 
way that the angle 
$i$ between the line of sight and the normal to the 
plane does not exceed $10^0$. In the catalog [16],
the fraction of 
such nuclei (among those with redshifts $z<0.01$) 
is $\sim 0.15$. However, the Seyfert nuclei 
identified as possible CR sources are 
oriented differently; the mean angle 
$i$ is $\approx 52^0$. If these objects 
are actually CR sources, then they emit 
uncollimated beams of particles. UHECR 
particles can be emitted in an uncollimated 
way from an accretion disk in the model 
[17]. We adopt 
this model by assuming that CRs emerging 
from the source in a cone with an angle 
$\theta\approx 50^0$ are detected on Earth.

At a given CR intensity near the Earth, 
the power of a source with emission in a 
solid angle $\Omega_S$ is a factor of 
$\Omega_S/(4\pi)$ lower than that of 
an isotropic source. For an angle 
$\theta\approx 50^0$, the solid 
angle is $\Omega_S\approx 2$, 
and the UHECR beam power is 
$L^b_S\approx 2\times 10^{39}\,\mbox{erg}\,\,\mbox{s}^{-1}$ 
if $I(E)\approx 10^{-39}\,(\mbox {cm}^2\,\mbox{s}\,\mbox{sr}\,\mbox{eV})^{-1}$ 
and $L_S\approx 10^{40}\,\mbox{erg}\,\,\mbox{s}^{-1}$. 
Assuming that the energy spectrum of the 
accelerated particles is a power law [4]
and, therefore, the 
overwhelming majority of particles 
has an energy $E_0\approx 5\times 10^{19}$ eV, 
we find that the source emits 
$N_S = L^b_S/E_0\approx 2\times 10^{31}\,(\mbox {particle}\,\,\mbox{s}^{-1})$. 
At distance $R$ from the source, 
the beam particles cross the area $\pi R^2\Omega_S$, 
and the particle flux (the number of particles 
crossing a unit area perpendicular to the beam 
axis per unit time) is $N=L^b_S/(R^2\Omega_SE_0$) 
(this relationship is formally identical to the 
expression for the number of particles in a 
unit solid angle for isotropic emission $N=L_S/(4\pi R^2E_0$)).

Let the source be at a distance of $R=16$ Mpc. 
This is the distance at which the maximum in 
the spatial distribution of Seyfert nuclei with 
redshifts $z < 0.01$ from the catalog [18]
is located for 
the Hubble constant $H =75\,\mbox {km}\,\mbox{Mpc}^{-1}\,\mbox{s}^{-1}$. 
On Earth, the particle flux from this source 
is $N\approx 4.3\times 10^{-21}\,(\mbox {cm}^2\,\mbox{s})^{-1}$, 
and arrays with areas $S\approx 10, 100$, 
and $3000\,\mbox {km}^2$ detect 0.012, 0.12, 
and $\sim 4$ particles per year, respectively. 
Hence, we find that no doublets of particles 
are detected from a single source on arrays 
with $S\sim 10\,\mbox {km}^2$. 
On the AGASA array ($S\approx 100\,\mbox {km}^2$), 
a doublet of particles from a single source 
can be detected in an
observing time of $T>10$ yr. On the Pierre 
Auger array ($S\sim 3000\,\mbox {km}^2$), a cluster from a single source 
can be detected during a year of operation. 
In the section "Discussion'' we list 
the reasons why these estimates can be considerably lower.

Let us now determine the detection rate if the 
UHECR sources are BL Lac objects. In these 
objects, the jets are directed toward the 
observer (here, we do not touch on uniformity 
or nonuniformity of the CR intensity distribution 
in different regions of the intergalactic space). 
Consider the possible values of the parameters 
that are necessary for our estimates by assuming 
that the particles are accelerated in the source 
in accordance with the model [10]. 

In the model [10], the power 
spent on UHECR emission is $L_{BL} = 1.5\times 10^{12}L_\odot$, 
where $L_\odot$ is the luminosity of the Sun. 
At $L\approx 3.86\times 10^{33}\,\mbox{erg}\,\,\mbox{s}^{-1}$, 
we have $L_{BL}\approx 6\times 10^{45}\,\mbox{erg}\,\,\mbox{s}^{-1}$. 
Since
the particle in this model are accelerated by an 
electric field, we assume the original CR spectrum 
to be monoenergetic. The energy of the accelerated 
protons in the model reaches $\sim 10^{20}-10^{21}$ 
eV [10, 11], 
in agreement with the maximum value of $10^{21}$ eV 
in the sources obtained in [19, 20] 
and with our estimates [13]. 
According to our estimates obtained by 
comparing the calculated and measured UHECR 
spectra [13], the CR emission 
power is $L_{BL}\sim 10^{43}-10^{44}\,\mbox{erg}\,\,\mbox{s}^{-1}$
(this quantity cannot be determined with 
a better accuracy because of the large measurement errors). 
The particles are emitted in a directed 
beam with an opening half-angle 
$\alpha\approx 7\times 10^{-7}$ [10].

The fraction of the particles that reached 
the array with energy $E\ge 4\times 10^{19}$ eV 
depends on the distance of the source due to the 
CR energy losses in the interactions with the 
background radiations in the intergalactic space 
[21, 22]. 
In what follows, we assume that this fraction 
is 0.3, 0.5, 0.7, and 1 for distances $R\approx 800$, 
700, 600 Mpc, and $R<600$ Mpc, respectively. We will 
disregard the contribution from the sources at $R>800$ 
Mpc (because of the low particle density, 
the contribution from these sources to 
the detected flux is an order of magnitude 
lower than the contribution from the sources 
at $R<600$ Mpc even without including the CR 
energy losses in the intergalactic space).

Assuming that the source has a redshift $z\approx 0.2$
(this is the redshift at which the maximum in the
spatial distribution of BL Lac objects from the catalog [18]is located),
i.e., is
$R\approx 600$ Mpc away, we estimate the CR detection
rate for the following sets of parameters.

(1) Assuming that 
$L_{BL}\approx 6\times 10^{45}\,\mbox{erg}\,\,\mbox{s}^{-1}$ is the
power of the directed beam, we find for $E_0 = 10^{21}$ eV
that an array with a relatively small area, 
$S\approx 10\,\mbox {km}^2$,
detects $\sim 10^8$ particles per year from a single source,
in conflict with the measurements.

(2) For an isotropic CR emission power 
$L_{BL}\approx 6\times 10^{45}\,\mbox{erg}\,\,\mbox{s}^{-1}$ 
(the equivalent power of the directed
beam is $L^b_{BL}\approx 10^{33}\,\mbox{erg}\,\,\mbox{s}^{-1}$)
and a particle energy of
$E_0=10^{21}$ eV, an array with an area 
$S\approx 100\,\mbox {km}^2$ will
detect 1.8 particles per year from 
a single source and a
cluster of three or more particles 
in several years. On
an array with $S\approx 10\,\mbox {km}^2$, 
a doublet of particles can
be detected in a time $T>10$ yr.

(3) Finally, let us consider a set of parameters with
an isotropic CR emission power of $L_{BL}\sim 10^{44}\,\mbox{erg}\,\,\mbox{s}^{-1}$ 
(the equivalent power of the directed beam 
is $L^b_{BL}\approx 1.6\times 10^{31}\,\mbox{erg}\,\,\mbox{s}^{-1}$) 
and a particle energy of $E_0 =10^{21}$ eV. 
In this case, arrays with an area of 10, 100,
and $3000\,\mbox{km}^2$ 
will detect $5\times10^{-4}$, $5\times 10^{-3}$, 
and 0.15
particles per year, respectively, from a single source.
In this model, at $L_{BL}\le 10^{44}\,\mbox{erg}\,\,\mbox{s}^{-1}$, CR clusters
from a single source can be detected only on an array
with an area $S\sim 3000\,\mbox {km}^2$ in a time $T>10$ yr.

Below, we analyze the second and third sets of
parameters and do not consider the first set, because
it is in conflict with the experimental data. Let us turn
to the AGASA data.

\vspace*{3mm}

{\bf 3.~~ Analysis of the clusters detected on the AGASA array}

\vspace*{3mm}

UHECR particles whose arrival directions coincided
within a single error, a total of five doublets and
one triplet of 63 UHE particles, were detected on the
AGASA array over 10 years [23].
A particle detected on the Yakutsk array [24] 
also falls into the triplet particle region.
The detection rate of particles in clusters is $\sim 1-1.5$
particles per year; in doublet C4 (according to the
numbering of clusters from [23]),
the particles were detected with an interval of almost
10 years.

The UHECR particles most likely arrive from
areas of the sky with an enhanced AGN density [3]. 
(The sources in the model with the
second set of parameters, as shown above, constitute
an exception.) If this is the case, then the particles
that form a cluster also arrive from such areas. The
size of these areas in equatorial coordinates is 
$(\Delta\alpha\le 9^0, \Delta\delta\le 9^0)$. If 
we assume that the errors in the right
ascension and declination are approximately $3^0$, then
this is the region of a triple error in the determination
of the particle arrival direction.

Let us test this assumption. Let us consider the
AGNs near the arrival direction of the particles in each
cluster and estimate the CR detection rate. The size of
the neighborhood in which we will analyze the AGNs
is $(\Delta\alpha\le 9^0, \Delta\delta\le 9^0)$.

Seyfert nuclei with $z<0.01$ fall into the search region
of four clusters: C2, C3, C5, and C6. The search
region of triplet C2 contains five nuclei from the catalog [18], 
eight objects
from [16], and ten nuclei
from the catalog [25]. One
object from the catalog [25] falls into the region of doublet C3 
and one object from [16, 18, 25] falls into the region of cluster C5. 
There are five, three,
and four nuclei from the catalogs [16, 18, 25],
respectively, in the region of doublet C6.

The UHECR flux from the Seyfert nuclei in the
region of triplet C2 detected by an array with an area
$S\approx 100\,\mbox {km}^2$ is $\sim 0.6$ 
particles per year, the flux in
the region of doublets C3 and C5 is $\sim 0.03$ particles
per year, and the flux from the nuclei in the region of
cluster C6 is $\sim 2$ particles per year, as estimated from
the data of the catalogs [16] or [25].

The fluxes of $\sim 0.03$ particles per year are too
low to explain the doublets of particles; the fluxes of
$\sim 1$ particles per year are enough for clusters to be
detected on the AGASA array. Thus, the model with
Seyfert nuclei explains the origin of two clusters.

BL Lac objects fall into the search regions of all the
clusters, except doublet C4. Nearby Seyfert nuclei do
not fall into the C4 search region either. This may be
attributable to the lowGalactic latitude of the particle
arrival, $b\approx -10^0$, since this latitude corresponds to
the zone of avoidance of galaxies in which relatively
fewobjects are observed. The numbers of objects from
the catalogs [16, 18]
that fall into the search regions are respectively 4 and
3 for doublet C1, 9 and 2 for triplet C2, 8 and 4 for
cluster C3, 12 and 6 for doublet C5, 13 and 8 for
cluster C6.

Let us determine the UHECR flux from the
BL Lac objects in the cluster region initially for the
third set of parameters. (In our estimates, we disregarded
the BL Lac objects with unknown redshifts.)
An array with an area $S\approx 100\,\mbox {km}^2$ 
will detect the
following fluxes: $\sim 0.3, \sim 0.4, \sim 0.1, \sim 0.6$, and $\sim 0.45$
particles per year in the regions of clusters C1, C2,
C3, C5, and C6, respectively. The fluxes $\sim 0.3-0.6$
particles per year are enough to explain the detected
clusters of particles.

The fluxes calculated with the second set of parameters
will be a factor of $\sim 50$ higher. In this model,
the clusters of particles are emitted by single sources.

\vspace*{3mm}

{\bf 4.~~Discussion}

\vspace*{3mm}

The above fluxes may have been overestimated.
The reasons are the following.

First, the fluxes were obtained by assuming that
the array detects the emission from the sources during
the entire period of its operation. In fact, the source
position in the sky can depend on the time of the day
and on the season. As a result, the actual time during
which the source emission is detected by the array can
be considerably shorter. For instance, if the sky area
from which the cluster particles arrive falls into the
array survey region $\sim 1/2$ day during $\sim 1/2$ 
year, then
the detected CR flux from the sources will be a factor
of 4 lower than the estimates obtained.

Second, the arrays usually select showers with polar
angles of the arrival direction $\Theta<30^0-45^0$; 
therefore,
no more than half of the emission from a given
source is detected. As a result, the estimated CR
detection rate may be a factor of $\sim 2-10$ higher than
the measured values. For these reasons, although the
second set of parameters in the BL Lac model yields a
high CR flux, it is probably suitable for explaining the
detected clusters.

In addition, the activity of BL Lac objects can be
not constant, but quasi-periodic, with a period of 4 to
$\sim 25$ yr with a relatively short duration of the active
state [26]. If this is the case, then
the clusters can be formed through the emission and
the subsequent ''turn-off'' of a single source in the
model with the second set of parameters. Further
evidence for this picture is that BL Lac objects with
intense emission are the most likely sources of UHECRs [6]. 
The relatively short
flares of the CR sources are also consistent with the
formation of clusters of particles by several sources in
the model with the third set of parameters.

In addition, the CR emission may be affected by
the content of protons (nuclei) in the jet plasma or
in the region of the accretion disk where the particles
are accelerated. According to [27, 28], the fraction of
the protons in a jet can be $\sim 0.01-0.1$. A change in
the proton fraction in the particle acceleration region
by several factors will lead to a change in the CR
intensity also by several factors, and this will appear
as the turn-on and turn-off of the UHECR source.

The turn-on and turn-off of the sources may be
the reason why the AGASA array has detected only
doublets and one triplet and no clusters with a larger
number of particles over 10 years of its operation.

Let us now list the predictions of the model [10] in
which the particles are emitted by BL Lac objects
with a maximum energy of $10^{27}$ eV.
For $L_{BL}\approx 6\times 10^{45}\,\mbox{erg}\,\,\mbox{s}^{-1}$, 
the CR fluxes in it are
too low: arrays with areas 
$S\sim 100$ and $1000\,\mbox {km}^2$
will detect no more than $10^{-7}$ and $10^{-6}$ 
particles per
year, respectively, from the cluster region. 
If the power of the directed beam is
$L_{BL}\approx 6\times 10^{45}\,\mbox{erg}\,\,\mbox{s}^{-1}$,
then the predicted fluxes are too high. Thus, for example,
an array with $S\sim 10\,\mbox {km}^2$ 
will detect $\sim 10^6$
particles per year from a single source, in conflict with
the measurements. The model with such parameters
does not describe the UHECR spectra measured on
different arrays [13] either.

It follows from this comparison that CR data can
be a test for some of the theoretical estimates pertaining
to AGNs.

\vspace*{3mm}

{\bf 5.~~Conclusions}

\vspace*{3mm}

The model in which the sources of UHECRs are
AGNs can explain the origin of the particle clusters
detected by AGASA. The clusters of particles arrive
from areas of the sky with an enhanced density of
these objects (the sizes of such sky areas in equatorial
coordinates are $\Delta\alpha < 9^0, \Delta\delta < 9^0$), 
but can also be
detected from individual sources. No clusters can be
detected on arrays with an area $S\sim 10 \,\mbox{km}^2$.

If the UHECR sources are Seyfert nuclei, then a
doublet of particles from a single Seyfert nucleus can
be detected on an array with an area $S\approx 100  \,\mbox{km}^2$
in an observing time of $T>10$ yr. On an array with
$S\sim 3000 \,\mbox{km}^2$, a cluster can be detected in a year of
its operation.

If UHECRs are emitted by BL Lac objects, then
a cluster of particles can be produced by a single
source with an emission power in the CR beam of
$\sim 10^{33}\,\mbox{erg}\,\,\mbox{s}^{-1}$. 
Doublets and triplets of particles from
such single sources can be detected by an array with
an area $S\approx 100\,\mbox{km}^2$ in $\sim 2-4$ yr. 
If the CR emission
power is $\sim 10^{31}\,\mbox{erg}\,\,\mbox{s}^{-1}$, then the clusters of
particles are emitted by a group of sources. Doublets
and triplets of particles will be detected by an array
with an area $S\approx 100 \,\mbox{km}^2$ also 
in $\sim 2-4$ yr of its
operation.

The CR emission and the cluster formation can
be a.ected by the following factors. First, CRs are
accelerated in the source quasi-periodically, with a
period of 4 to $\sim 25$ yr with a relatively 
short duration
of the active state, which is possible, as suggested
by the results [26]. Second,
the UHECR particle emission may be affected by the
changing (from $\sim 0.01$ to 0.1) fraction of the protons
(nuclei) in the jet plasma or in the particle acceleration
region. (Such a change in the proton fraction
is consistent with the results [28].)

The variable activity of the sources as well as the
decrease and increase in the proton fraction in a jet
appears as the turn-on and turn-off of the source.
This may be the reason why the AGASA array detected
only doublets and one triplet and no clusters
with a larger number of particles over 10 years of its
operation.

The estimates of the CR detection rate from which
these conclusions were drawn have been obtained
without allowance for the actual time during which
the array detected the emission from the sources.
In addition, we disregarded the selection of showers
by the polar angle of their arrival, that is why no
more than half of the emission from the source may
be detected. Therefore, a further study of clusters
requires taking into account the actual time of CR
detection and selecting showers by the polar angle of
their arrival.

A comparison of the CR data with theoretical estimates
can serve as a test for the models of AGNs, the
UHECR sources.

\vspace*{3mm}

{\bf Acknowledgments}

\vspace*{3mm}

I am grateful to A.V. Zasov and V.V. Lidskii
for discussions and to the referees for remarks.

\vspace*{3mm}

{\bf References}

\vspace*{3mm}



\vspace*{3mm}

\quad $[1]$\quad  V.S. Berezinsky et al., {\it Astrophysics of Cosmic Rays.} 
Ed. V.L Ginzburg. 

\qquad \quad Moscow, ''Nauka'', 1990.

\quad $[2]$\quad A.V. Uryson, JETP Lett. {\bf 64}, 77 (1996).

\quad $[3]$\quad A.V. Uryson, Astron. Zh. {\bf 78}, 686 (2001)  
[Astron. Rep. {\bf 45}, 591 (2001)].

\quad $[4]$\quad A.V. Uryson, Astron. Astrophys. Trans. {\bf 23}, 43 (2004).

\quad $[5]$\quad P.G. Tinyakov and I.I. Tkachev, JETP. Lett. {\bf 74}, 445 (2001).

\quad $[6]$\quad D.S. Gorbunov et. al., Astrophys. J. {\bf 577}, L93 (2002).

\quad $[7]$\quad P.P. Kronberg, Rep. Progr. Phys. {\bf 57}, 325 (1994).

\quad $[8]$\quad A.V. Uryson, Astron. Lett. {\bf 27}, 775 (2001).

\quad $[9]$\quad K. Dolag et al., JETP Lett. {\bf 79}, 583 (2004).

\quad $[10]$\quad N.S. Kardashev, Mon. Not. R. Astron. Soc. {\bf 276}, 515 (1995).

\quad $[11]$\quad A.A. Shatskii and N.S. Kardashev, Astron. Zh. {\bf 79}, 708, 2002.

\quad $[12]$\quad A.V. Uryson, Astron. Zh. {\bf 81}, 99 (2004) [Astron. Rep. {\bf 48}, 81 (2004)].

\quad $[13]$\quad A.V. Uryson, Astron. Lett. {\bf 30}, 816 (2004).

\quad $[14]$\quad M. Nagano and A.A. Watson, Rev.  Mod. Phys. {\bf 72}, 689 (2000).

\quad $[15]$\quad E.V. Derishev et al.,Phys. Rev. {\bf D68}, 043003 (2003).

\quad $[16]$\quad M.-P. Veron-Cetty and P. Veron, http://www.obshp. fr, 2003.

\quad $[17]$\quad C. A. Haswell et al.,Astrophys. J. {\bf 401}, 495 (1992).

\quad $[18]$\quad M.-P. Veron-Cetty and P. Veron, Astron. Astrophys.  {\bf 374}, 92 (2001).

\quad $[19]$\quad F. A. Aharonian et al.,Phys. Rev. {\bf D66}, 023005 (2002).

\quad $[20]$\quad M. V. Medvedev, Phys. Rev. {\bf E67}, 045401 (2003).

\quad $[21]$\quad K. Greisen, Phys. Rev. Lett. {\bf16}, 748 (1966).

\quad $[22]$\quad G. T. Zatsepin and V.A. Kuzmin, JETP Lett.{\bf 4}, 78 (1966).

\quad $[23]$\quad N. Hayashida et al., astroph/0008102, 2000.

\quad $[24]$\quad  B. N. Afanasiev et al., {\it Proceedings of International Symposium 

\qquad \quad ''Extremely High Energy Cosmic Rays: Astrophysics and Future Observatories''.} 

\qquad \quad\, Ed. M. Nagano. Inst. Cosmic-Ray Res., Tokyo, 1996.

\quad $[25]$\quad  V. A. Lipovetskii  et al., {\it Communications of the SAO.} No 55, 1987.

\quad $[26]$\quad T. B. Pyatunina et al., astroph/0502173, 2005.

\quad $[27]$\quad  S. A. Koryagin, {\it Private communication.} 2004.

\quad $[28]$\quad  V. V. Zheleznyakov and S. A. Koryagin, Astron. Lett. {\bf 28}, 809 (2002).

\end{document}